\documentstyle[12pt]{article}

\begin{document}

\begin{titlepage}

\title{Numerical simulations on the $4d$ Heisenberg spin glass}
\author{ Barbara COLUZZI }

\date{October 1994}

\maketitle

\begin{center}
Dipartimento di Fisica, Universit\`a di Roma
  {\em La Sapienza},\\
P. Aldo Moro 2,  00185 Roma,  Italy\\

\vspace{1truecm}
coluzzi@roma1.infn.it

\vspace{2truecm}

\begin{abstract}
We study the $4d$ Heisenberg spin glass model with Gaussian nearest-neighbor
interactions. We use finite size scaling to analyze the data. We find a
behavior consistent with a finite temperature spin glass transition. Our
estimates for the critical exponents agree with the results from
$\varepsilon$-expansion.
\end{abstract}

\end{center}

\vspace{4truecm}

\hspace{.1in} PACS Numbers 7510N-0270

\hspace{.1in} Preprint Roma1 $n^{o}$ 1054

\end{titlepage}

\vspace{3truecm}

\noindent
The lower critical dimension $d_{l}$ of the short range models
remains one of the most controversial questions in the spin glass
\cite{mpv}-\cite{by} theory. In the last ten years a consistent number of
works has been devoted to  studying the Ising model, which seems to be
\cite{i3, i3n} very close to  $d_{l}$ in $d=3$. In the case of the short range
isotropic Heisenberg spin glass, the conclusions of various computer
simulations in $d=3$ \cite{oys, mii}  agree that the system is below
$d_{l}$. Using domain wall renormalization group techniques, it was argued
\cite{sg, by} that $d_{l}=4$ for this model. To our knowledge, there are no
previous numerical simulations on the $4d$ Heisenberg spin glass apart from
an early work by Stauffer and Binder \cite{sb}, which  studied  the
time dependence of the Edward Anderson order parameter \cite{ea} $q_{EA}(t)$
for vector spin glasses from $d=2$ to $d=6$. They observed a change in the
behavior of the Heisenberg model in $d=4$ but concluded that probably there
was no finite $T$ phase transition.

We have studied the $4d$ isotropic Heisenberg spin glass model with Gaussian
nearest-neighbor interactions. We have simulated small lattices (with linear
size from $L=3$ to $L=5$), using finite size scaling to analyze the data. Our
main conclusion is that $d=4$ seems to be well above the $d_{l}$ of the model.

After defining the model and the quantities we have measured, we present
 a theoretical estimate for critical exponents obtained from
$\varepsilon$-expansion results \cite{green} and we discuss  the
computer simulations in some details. Finally,  we present  the
numerical results and the conclusions.

The Hamiltonian of the model is given by
\begin{equation}
{\cal H}[ \vec{\sigma} ]
\equiv - \frac{1}{2} \sum_{i,j=1}^{V} J_{ij} \vec{\sigma}_{i}
\vec{\sigma}_{j}
\label{ham}
\end{equation}
where the spins $\{ \vec{\sigma_{i}} \}$ belong to the unit three-dimensional
sphere. The sum runs over the nearest-neighbor pairs in a simple hypercubic
lattice with periodic boundary conditions and size $V=L^{4}$. The interactions
$J_{ij}$ are quenched independent random variables with a symmetric Gaussian
distribution
\begin{equation}
P(J_{ij}) = \frac{1}{\sqrt{2 \pi}} e^{\textstyle - J_{ij}^{2}/2}
\label{pdij}
\end{equation}

The spin glass correlation function can be defined as
\begin{equation}
G_{SG}(r) \equiv \frac{1}{2V} \sum_{| \vec{r}_{i}- \vec{r}_{j}|=r}
\overline{ < \vec{ \sigma}_{i} \vec{ \sigma}_{j}  >^{2} }
\end{equation}
where $< (\cdot ) >$ means thermodynamic average and $\overline{ ( \cdot ) }$
means average over samples.

In the thermodynamic limit, in the paramagnetic phase,
for distances $r >> \xi_{SG}$ we expect $G_{SG}(r) \propto e^{-r/ \xi_{SG}}$,
where we have introduced the spin glass correlation length $\xi_{SG}$.
The divergence  of $\xi_{SG}$ when the critical temperature $T_{c}$ is
approached,
\begin{equation}
\xi_{SG} \propto (T-T_{c})^{- \nu}
\label{dnu}
\end{equation}
is characterized by the critical exponent $\nu$. The power law decay of the
correlation function for large $r$ at the critical point,
\begin{equation}
G_{SG}(r) \propto \frac{1}{r^{d-2+ \eta}}
\label{deta}
\end{equation}
is described by the anomalous dimension $\eta$.

In order to understand the behavior of the model, it is interesting
to introduce two independent replicas of the system, with the same
disorder configuration. Since we are dealing with a vector spin glass, the
overlap is a second-rank tensor in spin space. We can define
\begin{equation}
q^{\mu \nu} \equiv \frac{1}{V} \sum_{i=1}^{V} \sigma^{\mu}_{i} \tau^{\nu}_{i}
\label{qmunu}
\end{equation}
where $\{ \vec{ \sigma} \}$ and $\{ \vec{ \tau } \}$ are the spins of the two
replicas and $\mu$, $\nu$ = 1, 3 refer to spin components.
In our case the system is completely isotropic, therefore we
expect to deal with one spin glass order parameter $q$. It is convenient
\cite{by} to consider the rotational invariant quantity
\begin{equation}
Q \equiv \sqrt{ \frac{1}{3} \sum_{\mu,\nu=1}^{3} (q^{\mu \nu})^{2} }
\label{qu}
\end{equation}

The order parameter probability distribution $P(q)$ is correspondingly given
by
\begin{equation}
P(q) \equiv \overline{ < \delta ( q - Q ) > }
\label{pdiq}
\end{equation}

One of the quantities we have calculated is the spin glass susceptibility,
defined as
\begin{equation}
\chi_{SG}(L,T) \equiv \frac{3}{V} \overline{ < \left ( \sum_{i=1}^{V}
\vec{\sigma_{i}} \vec{\tau_{i}} \right ) ^{2} >}
\label{sus}
\end{equation}
It can be easily verified that this definition is equivalent to
$\chi_{SG}~\equiv~3V~\overline{<q^{2}>}$, the factor 3 having been inserted to
obtain $\lim_{T \rightarrow \infty} \chi_{SG}(L,T) =1$, like in the Ising case.
In the thermodynamic limit we expect the spin glass susceptibility to diverge
when $T_{c}$ is approached from the paramagnetic phase as
\begin{equation}
\chi_{SG}(T) \propto (T - T_{c})^{- \gamma}
\label{dgamma}
\end{equation}
where, by hyperscaling, $\gamma = (2- \eta) \: \nu$.

If $T_{c}=0$ and the system is below $d_{l}$, we still expect
$\xi_{SG}(T)$ and $\chi_{SG}(T)$ to behave near the critical
point according to (\ref{dnu}) and (\ref{dgamma}) respectively, diverging with
power laws. We expect exponential divergences for $d \rightarrow d_{l}$,
because $\nu$, $\gamma \rightarrow \infty$ in this limit. In the $3d$ Ising
spin glass the difficulty in distinguishing between the system being at or
above  $d_{l}$ \cite{i3} is correlated to the large value of $\nu$
obtained from finite size scaling analysis. We will see that this seems not
to be  the case in our model.

The Binder parameter has proved very successful to establish the presence or
the absence of a finite $T$ phase transition. Its definition for the Ising
spin glass model \cite{bypq} can be  easily extended to the Heisenberg case.
In
the high temperature region, where we can neglect interactions, the
$q^{\mu \nu}$ are approximately independent variables with the same symmetric
Gaussian distribution of width $ \sim V^{-1/2}$. The function
\begin{equation}
g(L,T) \equiv \frac{1}{2} \left (11 -9 \: \frac{ \overline{ < q^{4} > }}
{ ( \overline{ <q^{2}>})^{2}} \right )
\label{pbin}
\end{equation}
is a dimensionless parameter defined so that $ g \leq 1$. From an explicit
calculation,  we have obtained $g \sim 1/V$ for  $T \rightarrow \infty$.
In the thermodynamic limit we expect therefore  $g(T)=0$ above $T_{c}$ and
$g(T)=1$ for $T=0$.
If there is a $T=0$ singularity we expect the  curves of $g(L,T)$ against $T$
for different $L$  to come together as $T \rightarrow 0$, while for a finite
$T$
transition we expect them to intersect at $T_{c}$, which allows to locate
the critical point quite precisely.

The mean field spin glass critical temperature $T_{c}^{MF}$ for $m$-component
spins belonging to the unit sphere, with coordination number $z$ and
$\overline{J_{ij}^{2}}=1$, is approximately, for large $z$,
$T_{c}^{MF} \simeq \sqrt{z}/m$. In our case $m=3$ and $z=8$, so
\begin{equation}
T_{c}^{MF} \simeq 0.94
\label{tcmf}
\end{equation}
This value is a reasonable upper limit to the temperature range in which we
can expect a phase transition.

If there is a nonzero $T_{c}$, the behavior of the system at the
transition point is characterized by two independent critical exponents. We
have obtained a theoretical estimate for the spin glass correlation length
exponent $\nu$ and the anomalous dimension $\eta$ from $\varepsilon$-expansion
results. This is possible in the Heisenberg case because the coefficients,
calculated to the third order by Green \cite{green}, show the expected
oscillatory behavior:
\begin{equation}
\begin{array}{rcl}
\eta & = & -0.2 \: \varepsilon + 7.7333 \cdot 10^{-2} \: \varepsilon^{2}
- 7.8127 \cdot 10^{-2} \: \varepsilon^{3} + O(\varepsilon^{4}) \\
\nu^{-1}-2+ \eta & = & -1.2 \: \varepsilon + 1.164 \: \varepsilon^{2}
- 1.4735 \: \varepsilon^{3} + O(\varepsilon^{4})
\end{array}
\end{equation}
where $\varepsilon=d_{u}-d=6-d$, $d_{u}=6$ being the upper critical dimension
of short range spin glass models.

We have used the simple resummation method of the Pad\'e approximants, in which
the series is replaced by the ratio ${\cal P}[N,D]$ of a polynomial of degree
$N$ to one of degree $D$. The values obtained for $\varepsilon=2$ are
presented in [Tab. 1]. By comparing results from ${\cal P}[1,2]$ and ${\cal
P}[2,1]$  we can estimate, for the $4d$ Heisenberg spin glass,
\begin{equation}
\nu \simeq 0.8 \hspace{.5in} \eta \simeq -0.4
\label{tene}
\end{equation}
with a larger uncertainty on the value of $\eta$.

\begin{table}
\begin{center}
\begin{tabular}{||c|c||c|c||}
\hline
\hline
\multicolumn{2}{||c||}{$\nu$} &  \multicolumn{2}{c||}{$\eta$} \\
\hline
\hline
0.7096 & 0.8657 & -0.2256 & -0.4945 \\
\hline
0.8224 & & -0.2976 &  \\
\hline
\hline
\end{tabular}
\caption{$\nu$ and $\eta$ from the corresponding ${\cal P}[N,D]$, $N$ being
the row and $D$ the column index.}
\end{center}
\end{table}

We have simulated hypercubic lattices in $4d$ with periodic boundary
conditions and linear sizes $L=3, 4$ and 5. The number of samples is 400, 200
and 100, respectively. Simulations have been performed in the region $T \leq
1$,
down to $T=0.4$ for $L=3$, to $T=0.45$ for $L=4$ and to $T=0.5$ for $L=5$.  We
will see that $T=0.5$ seems to be very close to the $T_{c}$ of our model.

In order to thermalize our samples, we have used Simulated Tempering
\cite{st1}-\cite{st3}, already proved very efficient for Ising spin glasses
\cite{i3n, i2}. The system, in our case consisting of two independent replicas,
is allowed to change temperature between a fixed set of $\{ \beta_{n} \}$,
where we can take for simplicity $\beta_{n+1}>\beta_{n}$, $\beta=1/T$ becoming
a dynamical variable. The stationary probability distribution for the
configuration $C$ at $\beta_{n}$ is given by
\begin{equation}
P_{\infty}(C,n) \propto e^{\textstyle -{\cal H}_{tot}[C,n]} \hspace{.5in}
{\cal H}_{tot}[C,n] \equiv \beta_{n}{\cal H}[C]-g_{n}
\end{equation}
where we
have defined the extended Hamiltonian ${\cal H}_{tot}[C,n]$, $\{g_{n} \}$
being a set of arbitrary numbers chosen $a~priori$. After reaching the
equilibrium, the system moves between the $\{ \beta_{n} \}$   remaining at the
equilibrium. In order to obtain the same stationary probability for all the
different temperatures we have to take $g_{n}= \beta_{n} F_{n}$, $F_{n}$ being
the total free energy at $\beta_{n}$. This means that, for $n'=n \pm 1$, at
the first order in $\Delta \beta=(\beta_{n'}-\beta_{n})$:
\begin{equation}
\Delta {\cal H}_{tot}= \Delta \beta {\cal H} - (g_{n'}-g_{n})
\simeq \Delta \beta \left ( {\cal H} - \frac{1}{2} ( < {\cal H}_{n'} > +
< {\cal H}_{n} > ) \right )
\end{equation}
where ${\cal H}$ is the instantaneous value and $< {\cal H}_{n} >$ is the
statistical expectation value  at $\beta_{n}$ of the total energy.

Our Simulated Tempering steps are just Monte Carlo steps at the end of which
the system is allowed to change temperature, the new  $\beta_{n}$  suggested
being $\beta_{n \pm 1}$ with equal probability. Obviously during the MC step
the two replicas evolve independently. Since we are dealing with a model with
continuous degrees of freedom, there is an arbitrary parameter in the
Metropolis algorithm, corresponding to the maximum rotation angle
$\theta_{max}$ permitted to single spins in one step. We have chosen
$\theta_{max}$ in order to obtain the acceptance as close to 1/2 as possible.

We have been careful in fixing the set of temperatures for the different
sizes. Two contiguous values of $\beta_{n}$ have to be as different as
possible to help in decorrelating without making too small the corresponding
transition probability. We have used as a basic criterion the condition that
there was a non-negligible overlap in the values of the energy computed at
contiguous $\beta_{n}$ for each sample. In our case this was verified by
choosing  equidistant  temperatures,
with $\Delta T=0.1$ for $L=3$ and $\Delta T=0.05$
for $L=4$ and 5. In order to perform simulations down to lower temperatures,
particularly in the $L=5$ case, it would be necessary to decrease $\Delta T$
with a considerable greater amount of computer time.

We have used a slow cooling procedure to take the system near the equilibrium
at the lowest temperature. Statistics were collected over the last part of the
about $ 3000 \: L \: \beta_{n}^{2} \: \:$ MC steps at each temperature, to
evaluate approximately the corresponding $< {\cal H}_{n} >$. The next about
70000 for $L=3$, 200000 for $L=4$ and 400000 for $L=5$ Simulated Tempering
steps were used to thermalize the system and to improve iteratively \cite{i2}
the estimates for the $\{ g_{n'}-g_{n} \}$. Finally, all the quantities we were
interested in were calculated in the last part of the Simulated Tempering
cycle, of about 140000 steps for $L=3$,  half a million steps for $L=4$
and more than a million steps for $L=5$.

In Simulated Tempering, $\:$ we can estimate $\:$ the
statistical $\:$ expectation  value
$<~{\cal O}_{n}~>$  of an observable ${\cal O}$ at $\beta_{n}$ as
\begin{equation}
< {\cal O}_{n} > = \frac{ 1}{{\cal N}f_{n}} \sum_{t} {\cal O} (t)
\delta_{\beta(t) \beta_{n}} \hspace{.5in}
f_{n} \equiv \frac{1}{{\cal N}} \sum_{t} \delta_{ \beta(t) \beta_{n}}
\end{equation}
where  we have defined the frequency $f_{n}$ observed at  $\beta_{n}$,
${\cal N}$ being the total number of steps. With the given choice for the
$\{ g_{n} \}$, we expect $f_{n} \simeq 1/n_{T}$, $n_{T}$ being the total
number of temperatures considered (in our case, $n_{T}=7$ for $L=3$ and
$n_{T}=8$ for $L=4$ and 5). The closeness of the $f_{n}$ observed at different
temperatures represents one of the main verifications that Simulated Tempering
works well. In the last part of the cycle we have obtained
$\max \{ f_{n} \} < 4 \min \{ f_{n} \}$ for each sample, the
$\overline{ f_{n} }$ being compatible with the expected values for the
different $n_{T}$. We have also checked how many $n_{e}$ times the system was
moving from one extreme to the other of the temperature range, obtaining for
each sample $n_{e}>100$ in the $L=3$ case and $n_{e}>200$ for $L=4$ and 5,
reasonably large values for the system really exploring the entire phase space.

In order to check thermalization, besides verifying that the $< q^{\mu \nu} >$
were compatible with zero for each sample, we have divided the last part of
the Simulated Tempering cycle in five equal intervals, checking that the
various computed quantities show no evident drifts. This was verified also
for the spin glass susceptibility (\ref{sus}) and the fourth moment of the
$P(q)$ (\ref{pdiq}) in the $L=5$ case.

The simulations have taken in all about 6 months of Dec3000-workstation.

Our numerical results for the spin glass susceptibility (\ref{sus}) and the
Binder parameter (\ref{pbin}) are presented in [Fig. 1] and [Fig. 2]
respectively. The behavior of the curves of $g(L,T)$ as a function of $T$ for
the different lattice sizes strongly suggests the presence of a
finite $T$ spin glass transition, the noncoincidence of the intersection
points being presumably due to systematic corrections to finite size scaling.
In order to confirm that curves really splay out below the critical point, as
expected in the presence of long range spin glass order, it would be obviously
preferable to obtain data on the $g(L,T)$ down to lower temperatures and for
larger lattice sizes.

If scaling is satisfied, near $T_{c}$  we expect
\begin{eqnarray}
\chi_{SG}(L,T)&=& L^{ 2-\eta}\tilde{\chi}_{SG}
\left ( (T-T_{c}) \: L^{ 1/ \nu} \right )
\label{susc}\\
g(L,T)&=& \tilde{g}\left ( (T-T_{c}) \: L^{ 1/ \nu} \right )
\label{bisc}
\end{eqnarray}
where $\tilde{\chi}_{SG}$ and $\tilde{g}$ are scaling functions while $\nu$
and $\eta$ are respectively the spin glass correlation length exponent
(\ref{dnu}) and the anomalous dimension (\ref{deta}) previously defined.

{}From the $\chi_{SG}$ scaling law (\ref{susc}), using a standard
three-parameter
fitting routine,  we have obtained
\begin{equation}
T_{c}= 0.50 \pm 0.06 \hspace{.3in}
\nu = 0.61 \pm 0.08 \hspace{.3in}
2 - \eta = 1.8 \pm 0.5
\label{schi}
\end{equation}
while requiring that the Binder parameter data scale according to equation
(\ref{bisc}), with a more simple two-parameter fit, we have found
\begin{equation}
T_{c}=0.52 \pm 0.02  \hspace{.5in}
\nu=0.89 \pm 0.06
\label{sbin}
\end{equation}
In [Fig. 3] and [Fig. 4] we present the corrispondent scaling plots. It must
be emphasized that statistical errors quoted here are just a delimitation of
the range of values beyond which our data do not scale well. Systematic errors
due to corrections to finite size scaling cannot easily be evaluated but could
be quite important, because of the small lattice sizes considered.

Our results agree well with a nonzero $T_{c}$ for the $4d$ short range
Heisenberg spin glass (\ref{ham}), giving in this case of Gaussian
nearest-neighbor interactions (\ref{pdij}) the value $T_{c} \simeq 0.5$, that
is
however well below the $T_{c}^{MF} \simeq 0.94$ (\ref{tcmf}) of the model.

Systematic corrections to finite size scaling may explain the discrepancy
between the estimates for $\nu$ obtained from the $\chi_{SG}$ (\ref{schi}) and
the Binder parameter (\ref{sbin}) respectively. Our most significant result is
nevertheless that $\nu$ seems to be not large, being both values smaller
than 1. As we have already pointed out, a large value of $\nu$ would suggest
the system being at $d \simeq d_{l}$, since we espect $\nu$,
$\gamma \rightarrow \infty$ for $d \rightarrow d_{l}$. In this case, therefore,
the result $\nu < 1$ is consistent with the short range Heisenberg spin glass
model being well above  $d_{l}$ in $d=4$.

Comparing the two estimates for $\nu$, we can obtain $\:$ the approximate
$\:$ value
$\nu \simeq 0.75$, that agrees well with the theoretical estimate
(\ref{tene}) $\nu \simeq 0.8$. From the hyperscaling law
$\alpha = 2 - d \: \nu$ we can also estimate the corresponding value of
 $\alpha$,  describing the critical behavior of the specific heat.
We find $\alpha \simeq -1$, accidentally not far from the mean field
value.

Finally, our statistics are inadequate to obtain a significant estimate for
$\eta$. The value found from the $\chi_{SG}$ data (\ref{schi}),
$2-\eta=1.8 \pm 0.5$, is however compatible with the theoretical estimate
(\ref{tene}) $\eta \simeq -0.4$, even if not in good agreement.

We have simulated small lattices, with a rather small amount of computer time.
More accurate simulations are necessary in order to confirm the presence
of a finite $T$ phase transition in the $4d$ short range Heisenberg spin glass
model, improving our estimates for critical exponent. Simulated Tempering
seems to be highly suitable for this purpose. It might be also interesting to
devote some attention to the behavior of the $P(q)$, to our knowledge not yet
extensively studied in the case of  short range vector models.

\vspace{.5cm}

I am deeply indebted to G. Parisi for his continuous support. I would also
like  to thank $\:$ E. Marinari for many helpful discussions, $\:$
A. Baldassarri, $\:$ F. Ritort and J. J. Ruiz for their useful suggestions.

\newpage

\begin{figure}
\setlength{\unitlength}{0.240900pt}
\ifx\plotpoint\undefined\newsavebox{\plotpoint}\fi
\sbox{\plotpoint}{\rule[-0.200pt]{0.400pt}{0.400pt}}%
\begin{picture}(1500,900)(0,0)
\font\gnuplot=cmr10 at 10pt
\gnuplot
\sbox{\plotpoint}{\rule[-0.200pt]{0.400pt}{0.400pt}}%
\put(220.0,113.0){\rule[-0.200pt]{4.818pt}{0.400pt}}
\put(198,113){\makebox(0,0)[r]{2}}
\put(1416.0,113.0){\rule[-0.200pt]{4.818pt}{0.400pt}}
\put(220.0,196.0){\rule[-0.200pt]{4.818pt}{0.400pt}}
\put(198,196){\makebox(0,0)[r]{4}}
\put(1416.0,196.0){\rule[-0.200pt]{4.818pt}{0.400pt}}
\put(220.0,278.0){\rule[-0.200pt]{4.818pt}{0.400pt}}
\put(198,278){\makebox(0,0)[r]{6}}
\put(1416.0,278.0){\rule[-0.200pt]{4.818pt}{0.400pt}}
\put(220.0,361.0){\rule[-0.200pt]{4.818pt}{0.400pt}}
\put(198,361){\makebox(0,0)[r]{8}}
\put(1416.0,361.0){\rule[-0.200pt]{4.818pt}{0.400pt}}
\put(220.0,443.0){\rule[-0.200pt]{4.818pt}{0.400pt}}
\put(198,443){\makebox(0,0)[r]{10}}
\put(1416.0,443.0){\rule[-0.200pt]{4.818pt}{0.400pt}}
\put(220.0,526.0){\rule[-0.200pt]{4.818pt}{0.400pt}}
\put(198,526){\makebox(0,0)[r]{12}}
\put(1416.0,526.0){\rule[-0.200pt]{4.818pt}{0.400pt}}
\put(220.0,609.0){\rule[-0.200pt]{4.818pt}{0.400pt}}
\put(198,609){\makebox(0,0)[r]{14}}
\put(1416.0,609.0){\rule[-0.200pt]{4.818pt}{0.400pt}}
\put(220.0,691.0){\rule[-0.200pt]{4.818pt}{0.400pt}}
\put(198,691){\makebox(0,0)[r]{16}}
\put(1416.0,691.0){\rule[-0.200pt]{4.818pt}{0.400pt}}
\put(220.0,774.0){\rule[-0.200pt]{4.818pt}{0.400pt}}
\put(198,774){\makebox(0,0)[r]{18}}
\put(1416.0,774.0){\rule[-0.200pt]{4.818pt}{0.400pt}}
\put(220.0,856.0){\rule[-0.200pt]{4.818pt}{0.400pt}}
\put(198,856){\makebox(0,0)[r]{20}}
\put(1416.0,856.0){\rule[-0.200pt]{4.818pt}{0.400pt}}
\put(292.0,113.0){\rule[-0.200pt]{0.400pt}{4.818pt}}
\put(292,68){\makebox(0,0){1}}
\put(292.0,857.0){\rule[-0.200pt]{0.400pt}{4.818pt}}
\put(435.0,113.0){\rule[-0.200pt]{0.400pt}{4.818pt}}
\put(435,68){\makebox(0,0){1.2}}
\put(435.0,857.0){\rule[-0.200pt]{0.400pt}{4.818pt}}
\put(578.0,113.0){\rule[-0.200pt]{0.400pt}{4.818pt}}
\put(578,68){\makebox(0,0){1.4}}
\put(578.0,857.0){\rule[-0.200pt]{0.400pt}{4.818pt}}
\put(721.0,113.0){\rule[-0.200pt]{0.400pt}{4.818pt}}
\put(721,68){\makebox(0,0){1.6}}
\put(721.0,857.0){\rule[-0.200pt]{0.400pt}{4.818pt}}
\put(864.0,113.0){\rule[-0.200pt]{0.400pt}{4.818pt}}
\put(864,68){\makebox(0,0){1.8}}
\put(864.0,857.0){\rule[-0.200pt]{0.400pt}{4.818pt}}
\put(1007.0,113.0){\rule[-0.200pt]{0.400pt}{4.818pt}}
\put(1007,68){\makebox(0,0){2}}
\put(1007.0,857.0){\rule[-0.200pt]{0.400pt}{4.818pt}}
\put(1150.0,113.0){\rule[-0.200pt]{0.400pt}{4.818pt}}
\put(1150,68){\makebox(0,0){2.2}}
\put(1150.0,857.0){\rule[-0.200pt]{0.400pt}{4.818pt}}
\put(1293.0,113.0){\rule[-0.200pt]{0.400pt}{4.818pt}}
\put(1293,68){\makebox(0,0){2.4}}
\put(1293.0,857.0){\rule[-0.200pt]{0.400pt}{4.818pt}}
\put(1436.0,113.0){\rule[-0.200pt]{0.400pt}{4.818pt}}
\put(1436,68){\makebox(0,0){2.6}}
\put(1436.0,857.0){\rule[-0.200pt]{0.400pt}{4.818pt}}
\put(220.0,113.0){\rule[-0.200pt]{292.934pt}{0.400pt}}
\put(1436.0,113.0){\rule[-0.200pt]{0.400pt}{184.048pt}}
\put(220.0,877.0){\rule[-0.200pt]{292.934pt}{0.400pt}}
\put(45,495){\makebox(0,0){$\chi_{SG}$}}
\put(828,23){\makebox(0,0){$\beta$}}
\put(220.0,113.0){\rule[-0.200pt]{0.400pt}{184.048pt}}
\put(292,125){\usebox{\plotpoint}}
\multiput(292.00,125.58)(2.362,0.495){31}{\rule{1.959pt}{0.119pt}}
\multiput(292.00,124.17)(74.934,17.000){2}{\rule{0.979pt}{0.400pt}}
\multiput(371.00,142.58)(1.850,0.497){51}{\rule{1.567pt}{0.120pt}}
\multiput(371.00,141.17)(95.748,27.000){2}{\rule{0.783pt}{0.400pt}}
\multiput(470.00,169.58)(1.650,0.498){75}{\rule{1.413pt}{0.120pt}}
\multiput(470.00,168.17)(125.068,39.000){2}{\rule{0.706pt}{0.400pt}}
\multiput(598.00,208.58)(1.446,0.499){115}{\rule{1.253pt}{0.120pt}}
\multiput(598.00,207.17)(167.400,59.000){2}{\rule{0.626pt}{0.400pt}}
\multiput(768.00,267.58)(1.409,0.499){167}{\rule{1.225pt}{0.120pt}}
\multiput(768.00,266.17)(236.458,85.000){2}{\rule{0.612pt}{0.400pt}}
\multiput(1007.00,352.58)(1.597,0.499){221}{\rule{1.375pt}{0.120pt}}
\multiput(1007.00,351.17)(354.146,112.000){2}{\rule{0.688pt}{0.400pt}}
\put(470,192){\usebox{\plotpoint}}
\multiput(470.00,192.58)(1.005,0.497){57}{\rule{0.900pt}{0.120pt}}
\multiput(470.00,191.17)(58.132,30.000){2}{\rule{0.450pt}{0.400pt}}
\multiput(530.00,222.58)(0.852,0.498){77}{\rule{0.780pt}{0.120pt}}
\multiput(530.00,221.17)(66.381,40.000){2}{\rule{0.390pt}{0.400pt}}
\multiput(598.00,262.58)(0.732,0.498){105}{\rule{0.685pt}{0.120pt}}
\multiput(598.00,261.17)(77.578,54.000){2}{\rule{0.343pt}{0.400pt}}
\multiput(677.00,316.58)(0.660,0.499){135}{\rule{0.628pt}{0.120pt}}
\multiput(677.00,315.17)(89.698,69.000){2}{\rule{0.314pt}{0.400pt}}
\multiput(768.00,385.58)(0.612,0.499){175}{\rule{0.590pt}{0.120pt}}
\multiput(768.00,384.17)(107.776,89.000){2}{\rule{0.295pt}{0.400pt}}
\multiput(877.00,474.58)(0.580,0.499){221}{\rule{0.564pt}{0.120pt}}
\multiput(877.00,473.17)(128.829,112.000){2}{\rule{0.282pt}{0.400pt}}
\multiput(1007.00,586.58)(0.580,0.499){271}{\rule{0.564pt}{0.120pt}}
\multiput(1007.00,585.17)(157.829,137.000){2}{\rule{0.282pt}{0.400pt}}
\put(418,174){\usebox{\plotpoint}}
\multiput(418.00,174.58)(1.046,0.497){47}{\rule{0.932pt}{0.120pt}}
\multiput(418.00,173.17)(50.066,25.000){2}{\rule{0.466pt}{0.400pt}}
\multiput(470.00,199.58)(0.836,0.498){69}{\rule{0.767pt}{0.120pt}}
\multiput(470.00,198.17)(58.409,36.000){2}{\rule{0.383pt}{0.400pt}}
\multiput(530.00,235.58)(0.667,0.498){99}{\rule{0.633pt}{0.120pt}}
\multiput(530.00,234.17)(66.685,51.000){2}{\rule{0.317pt}{0.400pt}}
\multiput(598.00,286.58)(0.533,0.499){145}{\rule{0.527pt}{0.120pt}}
\multiput(598.00,285.17)(77.906,74.000){2}{\rule{0.264pt}{0.400pt}}
\multiput(677.58,360.00)(0.499,0.588){179}{\rule{0.120pt}{0.570pt}}
\multiput(676.17,360.00)(91.000,105.816){2}{\rule{0.400pt}{0.285pt}}
\multiput(768.58,467.00)(0.499,0.698){215}{\rule{0.120pt}{0.658pt}}
\multiput(767.17,467.00)(109.000,150.635){2}{\rule{0.400pt}{0.329pt}}
\multiput(877.58,619.00)(0.499,0.808){257}{\rule{0.120pt}{0.746pt}}
\multiput(876.17,619.00)(130.000,208.451){2}{\rule{0.400pt}{0.373pt}}
\put(1306,767){\makebox(0,0)[r]{$L=3$}}
\put(1350,767){\raisebox{-.8pt}{\makebox(0,0){$\Diamond$}}}
\put(292,125){\raisebox{-.8pt}{\makebox(0,0){$\Diamond$}}}
\put(371,142){\raisebox{-.8pt}{\makebox(0,0){$\Diamond$}}}
\put(470,169){\raisebox{-.8pt}{\makebox(0,0){$\Diamond$}}}
\put(598,208){\raisebox{-.8pt}{\makebox(0,0){$\Diamond$}}}
\put(768,267){\raisebox{-.8pt}{\makebox(0,0){$\Diamond$}}}
\put(1007,352){\raisebox{-.8pt}{\makebox(0,0){$\Diamond$}}}
\put(1364,464){\raisebox{-.8pt}{\makebox(0,0){$\Diamond$}}}
\sbox{\plotpoint}{\rule[-0.400pt]{0.800pt}{0.800pt}}%
\put(1306,722){\makebox(0,0)[r]{$L=4$}}
\put(1350,722){\makebox(0,0){$+$}}
\put(470,192){\makebox(0,0){$+$}}
\put(530,222){\makebox(0,0){$+$}}
\put(598,262){\makebox(0,0){$+$}}
\put(677,316){\makebox(0,0){$+$}}
\put(768,385){\makebox(0,0){$+$}}
\put(877,474){\makebox(0,0){$+$}}
\put(1007,586){\makebox(0,0){$+$}}
\put(1166,723){\makebox(0,0){$+$}}
\sbox{\plotpoint}{\rule[-0.500pt]{1.000pt}{1.000pt}}%
\put(1306,677){\makebox(0,0)[r]{$L=5$}}
\put(1350,677){\raisebox{-.8pt}{\makebox(0,0){$\Box$}}}
\put(418,174){\raisebox{-.8pt}{\makebox(0,0){$\Box$}}}
\put(470,199){\raisebox{-.8pt}{\makebox(0,0){$\Box$}}}
\put(530,235){\raisebox{-.8pt}{\makebox(0,0){$\Box$}}}
\put(598,286){\raisebox{-.8pt}{\makebox(0,0){$\Box$}}}
\put(677,360){\raisebox{-.8pt}{\makebox(0,0){$\Box$}}}
\put(768,467){\raisebox{-.8pt}{\makebox(0,0){$\Box$}}}
\put(877,619){\raisebox{-.8pt}{\makebox(0,0){$\Box$}}}
\put(1007,829){\raisebox{-.8pt}{\makebox(0,0){$\Box$}}}
\end{picture}

\caption[]{The spin glass susceptibility $\chi_{SG}$ as a function of $\beta$
for the  different lattice sizes. Lines are only to join neighboring points.}
\label{fig:1}
\end{figure}
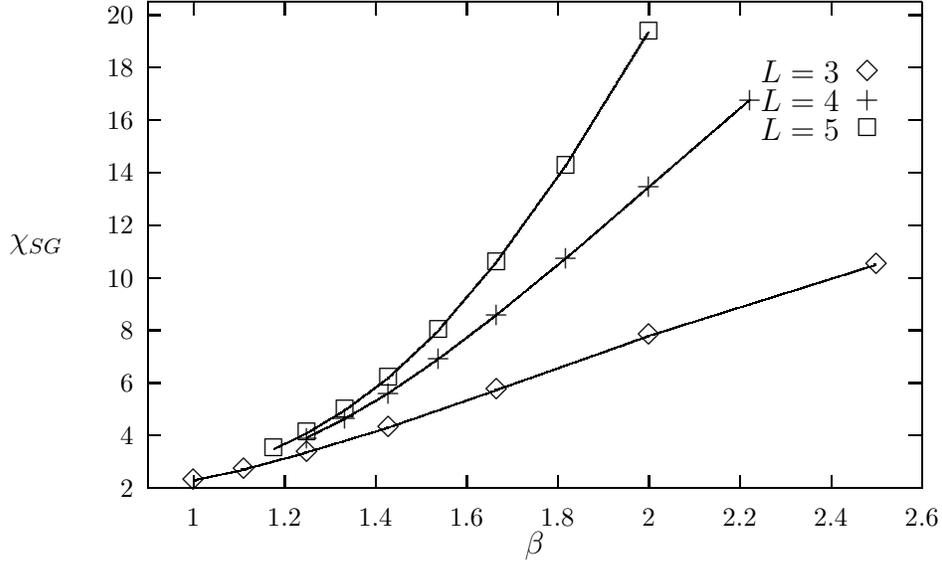

\begin{figure}

\setlength{\unitlength}{0.240900pt}
\ifx\plotpoint\undefined\newsavebox{\plotpoint}\fi
\sbox{\plotpoint}{\rule[-0.200pt]{0.400pt}{0.400pt}}%
\begin{picture}(1500,900)(0,0)
\font\gnuplot=cmr10 at 10pt
\gnuplot
\sbox{\plotpoint}{\rule[-0.200pt]{0.400pt}{0.400pt}}%
\put(220.0,113.0){\rule[-0.200pt]{292.934pt}{0.400pt}}
\put(220.0,113.0){\rule[-0.200pt]{4.818pt}{0.400pt}}
\put(198,113){\makebox(0,0)[r]{0}}
\put(1416.0,113.0){\rule[-0.200pt]{4.818pt}{0.400pt}}
\put(220.0,231.0){\rule[-0.200pt]{4.818pt}{0.400pt}}
\put(198,231){\makebox(0,0)[r]{0.1}}
\put(1416.0,231.0){\rule[-0.200pt]{4.818pt}{0.400pt}}
\put(220.0,348.0){\rule[-0.200pt]{4.818pt}{0.400pt}}
\put(198,348){\makebox(0,0)[r]{0.2}}
\put(1416.0,348.0){\rule[-0.200pt]{4.818pt}{0.400pt}}
\put(220.0,466.0){\rule[-0.200pt]{4.818pt}{0.400pt}}
\put(198,466){\makebox(0,0)[r]{0.3}}
\put(1416.0,466.0){\rule[-0.200pt]{4.818pt}{0.400pt}}
\put(220.0,583.0){\rule[-0.200pt]{4.818pt}{0.400pt}}
\put(198,583){\makebox(0,0)[r]{0.4}}
\put(1416.0,583.0){\rule[-0.200pt]{4.818pt}{0.400pt}}
\put(220.0,701.0){\rule[-0.200pt]{4.818pt}{0.400pt}}
\put(198,701){\makebox(0,0)[r]{0.5}}
\put(1416.0,701.0){\rule[-0.200pt]{4.818pt}{0.400pt}}
\put(220.0,818.0){\rule[-0.200pt]{4.818pt}{0.400pt}}
\put(198,818){\makebox(0,0)[r]{0.6}}
\put(1416.0,818.0){\rule[-0.200pt]{4.818pt}{0.400pt}}
\put(258.0,113.0){\rule[-0.200pt]{0.400pt}{4.818pt}}
\put(258,68){\makebox(0,0){0.4}}
\put(258.0,857.0){\rule[-0.200pt]{0.400pt}{4.818pt}}
\put(448.0,113.0){\rule[-0.200pt]{0.400pt}{4.818pt}}
\put(448,68){\makebox(0,0){0.5}}
\put(448.0,857.0){\rule[-0.200pt]{0.400pt}{4.818pt}}
\put(638.0,113.0){\rule[-0.200pt]{0.400pt}{4.818pt}}
\put(638,68){\makebox(0,0){0.6}}
\put(638.0,857.0){\rule[-0.200pt]{0.400pt}{4.818pt}}
\put(828.0,113.0){\rule[-0.200pt]{0.400pt}{4.818pt}}
\put(828,68){\makebox(0,0){0.7}}
\put(828.0,857.0){\rule[-0.200pt]{0.400pt}{4.818pt}}
\put(1018.0,113.0){\rule[-0.200pt]{0.400pt}{4.818pt}}
\put(1018,68){\makebox(0,0){0.8}}
\put(1018.0,857.0){\rule[-0.200pt]{0.400pt}{4.818pt}}
\put(1208.0,113.0){\rule[-0.200pt]{0.400pt}{4.818pt}}
\put(1208,68){\makebox(0,0){0.9}}
\put(1208.0,857.0){\rule[-0.200pt]{0.400pt}{4.818pt}}
\put(1398.0,113.0){\rule[-0.200pt]{0.400pt}{4.818pt}}
\put(1398,68){\makebox(0,0){1}}
\put(1398.0,857.0){\rule[-0.200pt]{0.400pt}{4.818pt}}
\put(220.0,113.0){\rule[-0.200pt]{292.934pt}{0.400pt}}
\put(1436.0,113.0){\rule[-0.200pt]{0.400pt}{184.048pt}}
\put(220.0,877.0){\rule[-0.200pt]{292.934pt}{0.400pt}}
\put(45,495){\makebox(0,0){$g$}}
\put(828,23){\makebox(0,0){$T$}}
\put(220.0,113.0){\rule[-0.200pt]{0.400pt}{184.048pt}}
\put(258,803){\usebox{\plotpoint}}
\multiput(258.58,800.85)(0.500,-0.524){377}{\rule{0.120pt}{0.519pt}}
\multiput(257.17,801.92)(190.000,-197.923){2}{\rule{0.400pt}{0.259pt}}
\multiput(448.00,602.92)(0.502,-0.500){375}{\rule{0.502pt}{0.120pt}}
\multiput(448.00,603.17)(188.958,-189.000){2}{\rule{0.251pt}{0.400pt}}
\multiput(638.00,413.92)(0.669,-0.499){281}{\rule{0.635pt}{0.120pt}}
\multiput(638.00,414.17)(188.682,-142.000){2}{\rule{0.318pt}{0.400pt}}
\multiput(828.00,271.92)(1.161,-0.499){161}{\rule{1.027pt}{0.120pt}}
\multiput(828.00,272.17)(187.869,-82.000){2}{\rule{0.513pt}{0.400pt}}
\multiput(1018.00,189.92)(2.587,-0.498){71}{\rule{2.154pt}{0.120pt}}
\multiput(1018.00,190.17)(185.529,-37.000){2}{\rule{1.077pt}{0.400pt}}
\multiput(1208.00,152.92)(6.067,-0.494){29}{\rule{4.850pt}{0.119pt}}
\multiput(1208.00,153.17)(179.934,-16.000){2}{\rule{2.425pt}{0.400pt}}
\put(353,749){\usebox{\plotpoint}}
\multiput(353.58,746.40)(0.499,-0.658){187}{\rule{0.120pt}{0.626pt}}
\multiput(352.17,747.70)(95.000,-123.700){2}{\rule{0.400pt}{0.313pt}}
\multiput(448.58,621.35)(0.499,-0.674){187}{\rule{0.120pt}{0.639pt}}
\multiput(447.17,622.67)(95.000,-126.674){2}{\rule{0.400pt}{0.319pt}}
\multiput(543.58,493.56)(0.499,-0.611){187}{\rule{0.120pt}{0.588pt}}
\multiput(542.17,494.78)(95.000,-114.779){2}{\rule{0.400pt}{0.294pt}}
\multiput(638.58,377.89)(0.499,-0.510){187}{\rule{0.120pt}{0.508pt}}
\multiput(637.17,378.94)(95.000,-95.945){2}{\rule{0.400pt}{0.254pt}}
\multiput(733.00,281.92)(0.699,-0.499){133}{\rule{0.659pt}{0.120pt}}
\multiput(733.00,282.17)(93.633,-68.000){2}{\rule{0.329pt}{0.400pt}}
\multiput(828.00,213.92)(1.059,-0.498){87}{\rule{0.944pt}{0.120pt}}
\multiput(828.00,214.17)(93.040,-45.000){2}{\rule{0.472pt}{0.400pt}}
\multiput(923.00,168.92)(1.711,-0.497){53}{\rule{1.457pt}{0.120pt}}
\multiput(923.00,169.17)(91.976,-28.000){2}{\rule{0.729pt}{0.400pt}}
\put(448,600){\usebox{\plotpoint}}
\multiput(448.58,597.05)(0.499,-0.764){187}{\rule{0.120pt}{0.711pt}}
\multiput(447.17,598.53)(95.000,-143.525){2}{\rule{0.400pt}{0.355pt}}
\multiput(543.58,452.42)(0.499,-0.653){187}{\rule{0.120pt}{0.622pt}}
\multiput(542.17,453.71)(95.000,-122.709){2}{\rule{0.400pt}{0.311pt}}
\multiput(638.58,328.89)(0.499,-0.510){187}{\rule{0.120pt}{0.508pt}}
\multiput(637.17,329.94)(95.000,-95.945){2}{\rule{0.400pt}{0.254pt}}
\multiput(733.00,232.92)(0.767,-0.499){121}{\rule{0.713pt}{0.120pt}}
\multiput(733.00,233.17)(93.520,-62.000){2}{\rule{0.356pt}{0.400pt}}
\multiput(828.00,170.92)(1.543,-0.497){59}{\rule{1.326pt}{0.120pt}}
\multiput(828.00,171.17)(92.248,-31.000){2}{\rule{0.663pt}{0.400pt}}
\multiput(923.00,139.92)(3.026,-0.494){29}{\rule{2.475pt}{0.119pt}}
\multiput(923.00,140.17)(89.863,-16.000){2}{\rule{1.237pt}{0.400pt}}
\multiput(1018.00,123.94)(13.787,-0.468){5}{\rule{9.600pt}{0.113pt}}
\multiput(1018.00,124.17)(75.075,-4.000){2}{\rule{4.800pt}{0.400pt}}
\put(1306,767){\makebox(0,0)[r]{$L=3$}}
\put(1350,767){\raisebox{-.8pt}{\makebox(0,0){$\Diamond$}}}
\put(258,803){\raisebox{-.8pt}{\makebox(0,0){$\Diamond$}}}
\put(448,604){\raisebox{-.8pt}{\makebox(0,0){$\Diamond$}}}
\put(638,415){\raisebox{-.8pt}{\makebox(0,0){$\Diamond$}}}
\put(828,273){\raisebox{-.8pt}{\makebox(0,0){$\Diamond$}}}
\put(1018,191){\raisebox{-.8pt}{\makebox(0,0){$\Diamond$}}}
\put(1208,154){\raisebox{-.8pt}{\makebox(0,0){$\Diamond$}}}
\put(1398,138){\raisebox{-.8pt}{\makebox(0,0){$\Diamond$}}}
\sbox{\plotpoint}{\rule[-0.400pt]{0.800pt}{0.800pt}}%
\put(1306,722){\makebox(0,0)[r]{$L=4$}}
\put(1350,722){\makebox(0,0){$+$}}
\put(353,749){\makebox(0,0){$+$}}
\put(448,624){\makebox(0,0){$+$}}
\put(543,496){\makebox(0,0){$+$}}
\put(638,380){\makebox(0,0){$+$}}
\put(733,283){\makebox(0,0){$+$}}
\put(828,215){\makebox(0,0){$+$}}
\put(923,170){\makebox(0,0){$+$}}
\put(1018,142){\makebox(0,0){$+$}}
\sbox{\plotpoint}{\rule[-0.500pt]{1.000pt}{1.000pt}}%
\put(1306,677){\makebox(0,0)[r]{$L=5$}}
\put(1350,677){\raisebox{-.8pt}{\makebox(0,0){$\Box$}}}
\put(448,600){\raisebox{-.8pt}{\makebox(0,0){$\Box$}}}
\put(543,455){\raisebox{-.8pt}{\makebox(0,0){$\Box$}}}
\put(638,331){\raisebox{-.8pt}{\makebox(0,0){$\Box$}}}
\put(733,234){\raisebox{-.8pt}{\makebox(0,0){$\Box$}}}
\put(828,172){\raisebox{-.8pt}{\makebox(0,0){$\Box$}}}
\put(923,141){\raisebox{-.8pt}{\makebox(0,0){$\Box$}}}
\put(1018,125){\raisebox{-.8pt}{\makebox(0,0){$\Box$}}}
\put(1113,121){\raisebox{-.8pt}{\makebox(0,0){$\Box$}}}
\end{picture}

\caption[]{The Binder parameter $g$ as a function of $T$ for the different
lattice sizes. Lines are only to join neighboring points.}
\label{fig:2}
\end{figure}
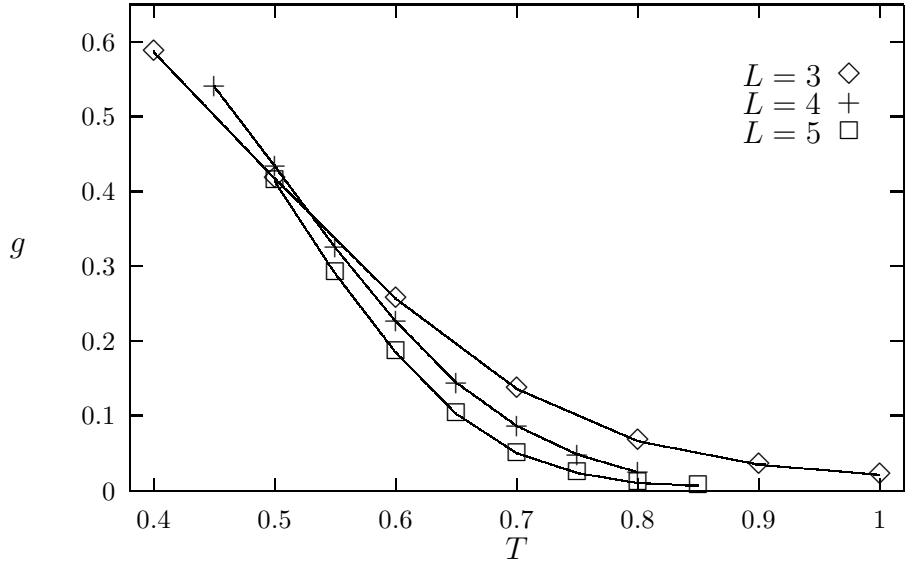

\begin{figure}

\setlength{\unitlength}{0.240900pt}
\ifx\plotpoint\undefined\newsavebox{\plotpoint}\fi
\sbox{\plotpoint}{\rule[-0.200pt]{0.400pt}{0.400pt}}%
\begin{picture}(1500,900)(0,0)
\font\gnuplot=cmr10 at 10pt
\gnuplot
\sbox{\plotpoint}{\rule[-0.200pt]{0.400pt}{0.400pt}}%
\put(220.0,113.0){\rule[-0.200pt]{4.818pt}{0.400pt}}
\put(198,113){\makebox(0,0)[r]{0.2}}
\put(1416.0,113.0){\rule[-0.200pt]{4.818pt}{0.400pt}}
\put(220.0,222.0){\rule[-0.200pt]{4.818pt}{0.400pt}}
\put(198,222){\makebox(0,0)[r]{0.4}}
\put(1416.0,222.0){\rule[-0.200pt]{4.818pt}{0.400pt}}
\put(220.0,331.0){\rule[-0.200pt]{4.818pt}{0.400pt}}
\put(198,331){\makebox(0,0)[r]{0.6}}
\put(1416.0,331.0){\rule[-0.200pt]{4.818pt}{0.400pt}}
\put(220.0,440.0){\rule[-0.200pt]{4.818pt}{0.400pt}}
\put(198,440){\makebox(0,0)[r]{0.8}}
\put(1416.0,440.0){\rule[-0.200pt]{4.818pt}{0.400pt}}
\put(220.0,550.0){\rule[-0.200pt]{4.818pt}{0.400pt}}
\put(198,550){\makebox(0,0)[r]{1}}
\put(1416.0,550.0){\rule[-0.200pt]{4.818pt}{0.400pt}}
\put(220.0,659.0){\rule[-0.200pt]{4.818pt}{0.400pt}}
\put(198,659){\makebox(0,0)[r]{1.2}}
\put(1416.0,659.0){\rule[-0.200pt]{4.818pt}{0.400pt}}
\put(220.0,768.0){\rule[-0.200pt]{4.818pt}{0.400pt}}
\put(198,768){\makebox(0,0)[r]{1.4}}
\put(1416.0,768.0){\rule[-0.200pt]{4.818pt}{0.400pt}}
\put(220.0,877.0){\rule[-0.200pt]{4.818pt}{0.400pt}}
\put(198,877){\makebox(0,0)[r]{1.6}}
\put(1416.0,877.0){\rule[-0.200pt]{4.818pt}{0.400pt}}
\put(220.0,113.0){\rule[-0.200pt]{0.400pt}{4.818pt}}
\put(220,68){\makebox(0,0){-1}}
\put(220.0,857.0){\rule[-0.200pt]{0.400pt}{4.818pt}}
\put(342.0,113.0){\rule[-0.200pt]{0.400pt}{4.818pt}}
\put(342,68){\makebox(0,0){-0.5}}
\put(342.0,857.0){\rule[-0.200pt]{0.400pt}{4.818pt}}
\put(463.0,113.0){\rule[-0.200pt]{0.400pt}{4.818pt}}
\put(463,68){\makebox(0,0){0}}
\put(463.0,857.0){\rule[-0.200pt]{0.400pt}{4.818pt}}
\put(585.0,113.0){\rule[-0.200pt]{0.400pt}{4.818pt}}
\put(585,68){\makebox(0,0){0.5}}
\put(585.0,857.0){\rule[-0.200pt]{0.400pt}{4.818pt}}
\put(706.0,113.0){\rule[-0.200pt]{0.400pt}{4.818pt}}
\put(706,68){\makebox(0,0){1}}
\put(706.0,857.0){\rule[-0.200pt]{0.400pt}{4.818pt}}
\put(828.0,113.0){\rule[-0.200pt]{0.400pt}{4.818pt}}
\put(828,68){\makebox(0,0){1.5}}
\put(828.0,857.0){\rule[-0.200pt]{0.400pt}{4.818pt}}
\put(950.0,113.0){\rule[-0.200pt]{0.400pt}{4.818pt}}
\put(950,68){\makebox(0,0){2}}
\put(950.0,857.0){\rule[-0.200pt]{0.400pt}{4.818pt}}
\put(1071.0,113.0){\rule[-0.200pt]{0.400pt}{4.818pt}}
\put(1071,68){\makebox(0,0){2.5}}
\put(1071.0,857.0){\rule[-0.200pt]{0.400pt}{4.818pt}}
\put(1193.0,113.0){\rule[-0.200pt]{0.400pt}{4.818pt}}
\put(1193,68){\makebox(0,0){3}}
\put(1193.0,857.0){\rule[-0.200pt]{0.400pt}{4.818pt}}
\put(1314.0,113.0){\rule[-0.200pt]{0.400pt}{4.818pt}}
\put(1314,68){\makebox(0,0){3.5}}
\put(1314.0,857.0){\rule[-0.200pt]{0.400pt}{4.818pt}}
\put(1436.0,113.0){\rule[-0.200pt]{0.400pt}{4.818pt}}
\put(1436,68){\makebox(0,0){4}}
\put(1436.0,857.0){\rule[-0.200pt]{0.400pt}{4.818pt}}
\put(220.0,113.0){\rule[-0.200pt]{292.934pt}{0.400pt}}
\put(1436.0,113.0){\rule[-0.200pt]{0.400pt}{184.048pt}}
\put(220.0,877.0){\rule[-0.200pt]{292.934pt}{0.400pt}}
\put(45,495){\makebox(0,0){$\chi_{SG}/L^{1.8}$}}
\put(828,23){\makebox(0,0){$(T-0.50)L^{1/0.62}$}}
\put(220.0,113.0){\rule[-0.200pt]{0.400pt}{184.048pt}}
\multiput(275.58,873.63)(0.495,-0.895){33}{\rule{0.119pt}{0.811pt}}
\multiput(274.17,875.32)(18.000,-30.316){2}{\rule{0.400pt}{0.406pt}}
\multiput(293.58,841.75)(0.496,-0.858){45}{\rule{0.120pt}{0.783pt}}
\multiput(292.17,843.37)(24.000,-39.374){2}{\rule{0.400pt}{0.392pt}}
\multiput(317.58,800.99)(0.497,-0.782){47}{\rule{0.120pt}{0.724pt}}
\multiput(316.17,802.50)(25.000,-37.497){2}{\rule{0.400pt}{0.362pt}}
\multiput(342.58,761.96)(0.496,-0.794){45}{\rule{0.120pt}{0.733pt}}
\multiput(341.17,763.48)(24.000,-36.478){2}{\rule{0.400pt}{0.367pt}}
\multiput(366.58,724.09)(0.496,-0.752){45}{\rule{0.120pt}{0.700pt}}
\multiput(365.17,725.55)(24.000,-34.547){2}{\rule{0.400pt}{0.350pt}}
\multiput(390.58,688.39)(0.497,-0.661){47}{\rule{0.120pt}{0.628pt}}
\multiput(389.17,689.70)(25.000,-31.697){2}{\rule{0.400pt}{0.314pt}}
\multiput(415.58,655.37)(0.496,-0.668){45}{\rule{0.120pt}{0.633pt}}
\multiput(414.17,656.69)(24.000,-30.685){2}{\rule{0.400pt}{0.317pt}}
\multiput(439.58,623.44)(0.496,-0.646){45}{\rule{0.120pt}{0.617pt}}
\multiput(438.17,624.72)(24.000,-29.720){2}{\rule{0.400pt}{0.308pt}}
\multiput(463.58,592.73)(0.497,-0.559){47}{\rule{0.120pt}{0.548pt}}
\multiput(462.17,593.86)(25.000,-26.863){2}{\rule{0.400pt}{0.274pt}}
\multiput(488.58,564.72)(0.496,-0.562){45}{\rule{0.120pt}{0.550pt}}
\multiput(487.17,565.86)(24.000,-25.858){2}{\rule{0.400pt}{0.275pt}}
\multiput(512.58,537.79)(0.496,-0.541){45}{\rule{0.120pt}{0.533pt}}
\multiput(511.17,538.89)(24.000,-24.893){2}{\rule{0.400pt}{0.267pt}}
\multiput(536.00,512.92)(0.498,-0.496){45}{\rule{0.500pt}{0.120pt}}
\multiput(536.00,513.17)(22.962,-24.000){2}{\rule{0.250pt}{0.400pt}}
\multiput(560.00,488.92)(0.542,-0.496){43}{\rule{0.535pt}{0.120pt}}
\multiput(560.00,489.17)(23.890,-23.000){2}{\rule{0.267pt}{0.400pt}}
\multiput(585.00,465.92)(0.544,-0.496){41}{\rule{0.536pt}{0.120pt}}
\multiput(585.00,466.17)(22.887,-22.000){2}{\rule{0.268pt}{0.400pt}}
\multiput(609.00,443.92)(0.600,-0.496){37}{\rule{0.580pt}{0.119pt}}
\multiput(609.00,444.17)(22.796,-20.000){2}{\rule{0.290pt}{0.400pt}}
\multiput(633.00,423.92)(0.625,-0.496){37}{\rule{0.600pt}{0.119pt}}
\multiput(633.00,424.17)(23.755,-20.000){2}{\rule{0.300pt}{0.400pt}}
\multiput(658.00,403.92)(0.668,-0.495){33}{\rule{0.633pt}{0.119pt}}
\multiput(658.00,404.17)(22.685,-18.000){2}{\rule{0.317pt}{0.400pt}}
\multiput(682.00,385.92)(0.708,-0.495){31}{\rule{0.665pt}{0.119pt}}
\multiput(682.00,386.17)(22.620,-17.000){2}{\rule{0.332pt}{0.400pt}}
\multiput(706.00,368.92)(0.785,-0.494){29}{\rule{0.725pt}{0.119pt}}
\multiput(706.00,369.17)(23.495,-16.000){2}{\rule{0.363pt}{0.400pt}}
\multiput(731.00,352.92)(0.805,-0.494){27}{\rule{0.740pt}{0.119pt}}
\multiput(731.00,353.17)(22.464,-15.000){2}{\rule{0.370pt}{0.400pt}}
\multiput(755.00,337.92)(0.864,-0.494){25}{\rule{0.786pt}{0.119pt}}
\multiput(755.00,338.17)(22.369,-14.000){2}{\rule{0.393pt}{0.400pt}}
\multiput(779.00,323.92)(0.901,-0.494){25}{\rule{0.814pt}{0.119pt}}
\multiput(779.00,324.17)(23.310,-14.000){2}{\rule{0.407pt}{0.400pt}}
\multiput(804.00,309.92)(1.013,-0.492){21}{\rule{0.900pt}{0.119pt}}
\multiput(804.00,310.17)(22.132,-12.000){2}{\rule{0.450pt}{0.400pt}}
\multiput(828.00,297.92)(1.013,-0.492){21}{\rule{0.900pt}{0.119pt}}
\multiput(828.00,298.17)(22.132,-12.000){2}{\rule{0.450pt}{0.400pt}}
\multiput(852.00,285.92)(1.156,-0.492){19}{\rule{1.009pt}{0.118pt}}
\multiput(852.00,286.17)(22.906,-11.000){2}{\rule{0.505pt}{0.400pt}}
\multiput(877.00,274.92)(1.109,-0.492){19}{\rule{0.973pt}{0.118pt}}
\multiput(877.00,275.17)(21.981,-11.000){2}{\rule{0.486pt}{0.400pt}}
\multiput(901.00,263.92)(1.225,-0.491){17}{\rule{1.060pt}{0.118pt}}
\multiput(901.00,264.17)(21.800,-10.000){2}{\rule{0.530pt}{0.400pt}}
\multiput(925.00,253.93)(1.427,-0.489){15}{\rule{1.211pt}{0.118pt}}
\multiput(925.00,254.17)(22.486,-9.000){2}{\rule{0.606pt}{0.400pt}}
\multiput(950.00,244.93)(1.368,-0.489){15}{\rule{1.167pt}{0.118pt}}
\multiput(950.00,245.17)(21.579,-9.000){2}{\rule{0.583pt}{0.400pt}}
\multiput(974.00,235.93)(1.550,-0.488){13}{\rule{1.300pt}{0.117pt}}
\multiput(974.00,236.17)(21.302,-8.000){2}{\rule{0.650pt}{0.400pt}}
\multiput(998.00,227.93)(1.616,-0.488){13}{\rule{1.350pt}{0.117pt}}
\multiput(998.00,228.17)(22.198,-8.000){2}{\rule{0.675pt}{0.400pt}}
\multiput(1023.00,219.93)(1.789,-0.485){11}{\rule{1.471pt}{0.117pt}}
\multiput(1023.00,220.17)(20.946,-7.000){2}{\rule{0.736pt}{0.400pt}}
\multiput(1047.00,212.93)(1.789,-0.485){11}{\rule{1.471pt}{0.117pt}}
\multiput(1047.00,213.17)(20.946,-7.000){2}{\rule{0.736pt}{0.400pt}}
\multiput(1071.00,205.93)(1.865,-0.485){11}{\rule{1.529pt}{0.117pt}}
\multiput(1071.00,206.17)(21.827,-7.000){2}{\rule{0.764pt}{0.400pt}}
\multiput(1096.00,198.93)(2.118,-0.482){9}{\rule{1.700pt}{0.116pt}}
\multiput(1096.00,199.17)(20.472,-6.000){2}{\rule{0.850pt}{0.400pt}}
\multiput(1120.00,192.93)(2.118,-0.482){9}{\rule{1.700pt}{0.116pt}}
\multiput(1120.00,193.17)(20.472,-6.000){2}{\rule{0.850pt}{0.400pt}}
\multiput(1144.00,186.93)(2.118,-0.482){9}{\rule{1.700pt}{0.116pt}}
\multiput(1144.00,187.17)(20.472,-6.000){2}{\rule{0.850pt}{0.400pt}}
\multiput(1168.00,180.93)(2.714,-0.477){7}{\rule{2.100pt}{0.115pt}}
\multiput(1168.00,181.17)(20.641,-5.000){2}{\rule{1.050pt}{0.400pt}}
\multiput(1193.00,175.93)(2.602,-0.477){7}{\rule{2.020pt}{0.115pt}}
\multiput(1193.00,176.17)(19.807,-5.000){2}{\rule{1.010pt}{0.400pt}}
\multiput(1217.00,170.93)(2.602,-0.477){7}{\rule{2.020pt}{0.115pt}}
\multiput(1217.00,171.17)(19.807,-5.000){2}{\rule{1.010pt}{0.400pt}}
\multiput(1241.00,165.93)(2.714,-0.477){7}{\rule{2.100pt}{0.115pt}}
\multiput(1241.00,166.17)(20.641,-5.000){2}{\rule{1.050pt}{0.400pt}}
\multiput(1266.00,160.94)(3.406,-0.468){5}{\rule{2.500pt}{0.113pt}}
\multiput(1266.00,161.17)(18.811,-4.000){2}{\rule{1.250pt}{0.400pt}}
\multiput(1290.00,156.94)(3.406,-0.468){5}{\rule{2.500pt}{0.113pt}}
\multiput(1290.00,157.17)(18.811,-4.000){2}{\rule{1.250pt}{0.400pt}}
\multiput(1314.00,152.93)(2.714,-0.477){7}{\rule{2.100pt}{0.115pt}}
\multiput(1314.00,153.17)(20.641,-5.000){2}{\rule{1.050pt}{0.400pt}}
\multiput(1339.00,147.94)(3.406,-0.468){5}{\rule{2.500pt}{0.113pt}}
\multiput(1339.00,148.17)(18.811,-4.000){2}{\rule{1.250pt}{0.400pt}}
\multiput(1363.00,143.95)(5.151,-0.447){3}{\rule{3.300pt}{0.108pt}}
\multiput(1363.00,144.17)(17.151,-3.000){2}{\rule{1.650pt}{0.400pt}}
\multiput(1387.00,140.94)(3.552,-0.468){5}{\rule{2.600pt}{0.113pt}}
\multiput(1387.00,141.17)(19.604,-4.000){2}{\rule{1.300pt}{0.400pt}}
\multiput(1412.00,136.94)(3.406,-0.468){5}{\rule{2.500pt}{0.113pt}}
\multiput(1412.00,137.17)(18.811,-4.000){2}{\rule{1.250pt}{0.400pt}}
\put(1306,767){\makebox(0,0)[r]{$L=3$}}
\put(1350,767){\raisebox{-.8pt}{\makebox(0,0){$\Diamond$}}}
\put(316,798){\raisebox{-.8pt}{\makebox(0,0){$\Diamond$}}}
\put(463,592){\raisebox{-.8pt}{\makebox(0,0){$\Diamond$}}}
\put(610,437){\raisebox{-.8pt}{\makebox(0,0){$\Diamond$}}}
\put(758,329){\raisebox{-.8pt}{\makebox(0,0){$\Diamond$}}}
\put(905,257){\raisebox{-.8pt}{\makebox(0,0){$\Diamond$}}}
\put(1052,209){\raisebox{-.8pt}{\makebox(0,0){$\Diamond$}}}
\put(1200,177){\raisebox{-.8pt}{\makebox(0,0){$\Diamond$}}}
\sbox{\plotpoint}{\rule[-0.400pt]{0.800pt}{0.800pt}}%
\put(1306,722){\makebox(0,0)[r]{$L=4$}}
\put(1350,722){\makebox(0,0){$+$}}
\put(345,759){\makebox(0,0){$+$}}
\put(463,610){\makebox(0,0){$+$}}
\put(581,487){\makebox(0,0){$+$}}
\put(699,390){\makebox(0,0){$+$}}
\put(817,315){\makebox(0,0){$+$}}
\put(935,256){\makebox(0,0){$+$}}
\put(1053,213){\makebox(0,0){$+$}}
\put(1171,180){\makebox(0,0){$+$}}
\sbox{\plotpoint}{\rule[-0.500pt]{1.000pt}{1.000pt}}%
\put(1306,677){\makebox(0,0)[r]{$L=5$}}
\put(1350,677){\raisebox{-.8pt}{\makebox(0,0){$\Box$}}}
\put(463,586){\raisebox{-.8pt}{\makebox(0,0){$\Box$}}}
\put(633,433){\raisebox{-.8pt}{\makebox(0,0){$\Box$}}}
\put(803,322){\raisebox{-.8pt}{\makebox(0,0){$\Box$}}}
\put(974,244){\raisebox{-.8pt}{\makebox(0,0){$\Box$}}}
\put(1144,190){\raisebox{-.8pt}{\makebox(0,0){$\Box$}}}
\put(1314,153){\raisebox{-.8pt}{\makebox(0,0){$\Box$}}}
\end{picture}

\caption[]{The scaled spin  glass  susceptibility $\chi_{SG} / L^{2-\eta}$  as
a
function of the scaled  reduced temperature $(T - T_{c}) L^{1/ \nu}$. The line
is just guide to the eye.}
\label{fig:3}
\end{figure}
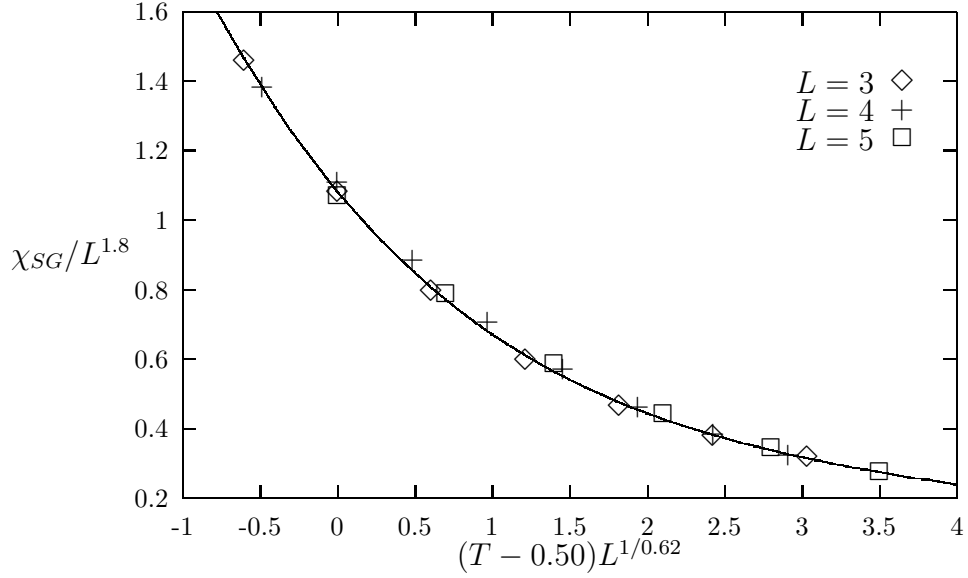

\begin{figure}
\setlength{\unitlength}{0.240900pt}
\ifx\plotpoint\undefined\newsavebox{\plotpoint}\fi
\sbox{\plotpoint}{\rule[-0.200pt]{0.400pt}{0.400pt}}%
\begin{picture}(1500,900)(0,0)
\font\gnuplot=cmr10 at 10pt
\gnuplot
\sbox{\plotpoint}{\rule[-0.200pt]{0.400pt}{0.400pt}}%
\put(220.0,113.0){\rule[-0.200pt]{4.818pt}{0.400pt}}
\put(198,113){\makebox(0,0)[r]{0}}
\put(1416.0,113.0){\rule[-0.200pt]{4.818pt}{0.400pt}}
\put(220.0,231.0){\rule[-0.200pt]{4.818pt}{0.400pt}}
\put(198,231){\makebox(0,0)[r]{0.1}}
\put(1416.0,231.0){\rule[-0.200pt]{4.818pt}{0.400pt}}
\put(220.0,348.0){\rule[-0.200pt]{4.818pt}{0.400pt}}
\put(198,348){\makebox(0,0)[r]{0.2}}
\put(1416.0,348.0){\rule[-0.200pt]{4.818pt}{0.400pt}}
\put(220.0,466.0){\rule[-0.200pt]{4.818pt}{0.400pt}}
\put(198,466){\makebox(0,0)[r]{0.3}}
\put(1416.0,466.0){\rule[-0.200pt]{4.818pt}{0.400pt}}
\put(220.0,583.0){\rule[-0.200pt]{4.818pt}{0.400pt}}
\put(198,583){\makebox(0,0)[r]{0.4}}
\put(1416.0,583.0){\rule[-0.200pt]{4.818pt}{0.400pt}}
\put(220.0,701.0){\rule[-0.200pt]{4.818pt}{0.400pt}}
\put(198,701){\makebox(0,0)[r]{0.5}}
\put(1416.0,701.0){\rule[-0.200pt]{4.818pt}{0.400pt}}
\put(220.0,818.0){\rule[-0.200pt]{4.818pt}{0.400pt}}
\put(198,818){\makebox(0,0)[r]{0.6}}
\put(1416.0,818.0){\rule[-0.200pt]{4.818pt}{0.400pt}}
\put(220.0,113.0){\rule[-0.200pt]{0.400pt}{4.818pt}}
\put(220,68){\makebox(0,0){-0.5}}
\put(220.0,857.0){\rule[-0.200pt]{0.400pt}{4.818pt}}
\put(484.0,113.0){\rule[-0.200pt]{0.400pt}{4.818pt}}
\put(484,68){\makebox(0,0){0}}
\put(484.0,857.0){\rule[-0.200pt]{0.400pt}{4.818pt}}
\put(749.0,113.0){\rule[-0.200pt]{0.400pt}{4.818pt}}
\put(749,68){\makebox(0,0){0.5}}
\put(749.0,857.0){\rule[-0.200pt]{0.400pt}{4.818pt}}
\put(1013.0,113.0){\rule[-0.200pt]{0.400pt}{4.818pt}}
\put(1013,68){\makebox(0,0){1}}
\put(1013.0,857.0){\rule[-0.200pt]{0.400pt}{4.818pt}}
\put(1277.0,113.0){\rule[-0.200pt]{0.400pt}{4.818pt}}
\put(1277,68){\makebox(0,0){1.5}}
\put(1277.0,857.0){\rule[-0.200pt]{0.400pt}{4.818pt}}
\put(220.0,113.0){\rule[-0.200pt]{292.934pt}{0.400pt}}
\put(1436.0,113.0){\rule[-0.200pt]{0.400pt}{184.048pt}}
\put(220.0,877.0){\rule[-0.200pt]{292.934pt}{0.400pt}}
\put(45,495){\makebox(0,0){$g$}}
\put(828,23){\makebox(0,0){$(T-0.52)L^{1/0.89}$}}
\put(220.0,113.0){\rule[-0.200pt]{0.400pt}{184.048pt}}
\put(220,870){\usebox{\plotpoint}}
\multiput(220.58,867.35)(0.497,-0.674){49}{\rule{0.120pt}{0.638pt}}
\multiput(219.17,868.67)(26.000,-33.675){2}{\rule{0.400pt}{0.319pt}}
\multiput(246.58,832.49)(0.497,-0.630){51}{\rule{0.120pt}{0.604pt}}
\multiput(245.17,833.75)(27.000,-32.747){2}{\rule{0.400pt}{0.302pt}}
\multiput(273.58,798.48)(0.497,-0.635){49}{\rule{0.120pt}{0.608pt}}
\multiput(272.17,799.74)(26.000,-31.739){2}{\rule{0.400pt}{0.304pt}}
\multiput(299.58,765.56)(0.497,-0.611){51}{\rule{0.120pt}{0.589pt}}
\multiput(298.17,766.78)(27.000,-31.778){2}{\rule{0.400pt}{0.294pt}}
\multiput(326.58,732.54)(0.497,-0.615){49}{\rule{0.120pt}{0.592pt}}
\multiput(325.17,733.77)(26.000,-30.771){2}{\rule{0.400pt}{0.296pt}}
\multiput(352.58,700.62)(0.497,-0.592){51}{\rule{0.120pt}{0.574pt}}
\multiput(351.17,701.81)(27.000,-30.808){2}{\rule{0.400pt}{0.287pt}}
\multiput(379.58,668.61)(0.497,-0.596){49}{\rule{0.120pt}{0.577pt}}
\multiput(378.17,669.80)(26.000,-29.803){2}{\rule{0.400pt}{0.288pt}}
\multiput(405.58,637.67)(0.497,-0.576){49}{\rule{0.120pt}{0.562pt}}
\multiput(404.17,638.83)(26.000,-28.834){2}{\rule{0.400pt}{0.281pt}}
\multiput(431.58,607.80)(0.497,-0.536){51}{\rule{0.120pt}{0.530pt}}
\multiput(430.17,608.90)(27.000,-27.901){2}{\rule{0.400pt}{0.265pt}}
\multiput(458.58,578.80)(0.497,-0.537){49}{\rule{0.120pt}{0.531pt}}
\multiput(457.17,579.90)(26.000,-26.898){2}{\rule{0.400pt}{0.265pt}}
\multiput(484.58,550.86)(0.497,-0.517){51}{\rule{0.120pt}{0.515pt}}
\multiput(483.17,551.93)(27.000,-26.931){2}{\rule{0.400pt}{0.257pt}}
\multiput(511.00,523.92)(0.498,-0.497){49}{\rule{0.500pt}{0.120pt}}
\multiput(511.00,524.17)(24.962,-26.000){2}{\rule{0.250pt}{0.400pt}}
\multiput(537.00,497.92)(0.518,-0.497){49}{\rule{0.515pt}{0.120pt}}
\multiput(537.00,498.17)(25.930,-26.000){2}{\rule{0.258pt}{0.400pt}}
\multiput(564.00,471.92)(0.541,-0.496){45}{\rule{0.533pt}{0.120pt}}
\multiput(564.00,472.17)(24.893,-24.000){2}{\rule{0.267pt}{0.400pt}}
\multiput(590.00,447.92)(0.562,-0.496){45}{\rule{0.550pt}{0.120pt}}
\multiput(590.00,448.17)(25.858,-24.000){2}{\rule{0.275pt}{0.400pt}}
\multiput(617.00,423.92)(0.564,-0.496){43}{\rule{0.552pt}{0.120pt}}
\multiput(617.00,424.17)(24.854,-23.000){2}{\rule{0.276pt}{0.400pt}}
\multiput(643.00,400.92)(0.591,-0.496){41}{\rule{0.573pt}{0.120pt}}
\multiput(643.00,401.17)(24.811,-22.000){2}{\rule{0.286pt}{0.400pt}}
\multiput(669.00,378.92)(0.676,-0.496){37}{\rule{0.640pt}{0.119pt}}
\multiput(669.00,379.17)(25.672,-20.000){2}{\rule{0.320pt}{0.400pt}}
\multiput(696.00,358.92)(0.651,-0.496){37}{\rule{0.620pt}{0.119pt}}
\multiput(696.00,359.17)(24.713,-20.000){2}{\rule{0.310pt}{0.400pt}}
\multiput(722.00,338.92)(0.713,-0.495){35}{\rule{0.668pt}{0.119pt}}
\multiput(722.00,339.17)(25.613,-19.000){2}{\rule{0.334pt}{0.400pt}}
\multiput(749.00,319.92)(0.725,-0.495){33}{\rule{0.678pt}{0.119pt}}
\multiput(749.00,320.17)(24.593,-18.000){2}{\rule{0.339pt}{0.400pt}}
\multiput(775.00,301.92)(0.849,-0.494){29}{\rule{0.775pt}{0.119pt}}
\multiput(775.00,302.17)(25.391,-16.000){2}{\rule{0.388pt}{0.400pt}}
\multiput(802.00,285.92)(0.817,-0.494){29}{\rule{0.750pt}{0.119pt}}
\multiput(802.00,286.17)(24.443,-16.000){2}{\rule{0.375pt}{0.400pt}}
\multiput(828.00,269.92)(0.873,-0.494){27}{\rule{0.793pt}{0.119pt}}
\multiput(828.00,270.17)(24.353,-15.000){2}{\rule{0.397pt}{0.400pt}}
\multiput(854.00,254.92)(0.974,-0.494){25}{\rule{0.871pt}{0.119pt}}
\multiput(854.00,255.17)(25.191,-14.000){2}{\rule{0.436pt}{0.400pt}}
\multiput(881.00,240.92)(1.099,-0.492){21}{\rule{0.967pt}{0.119pt}}
\multiput(881.00,241.17)(23.994,-12.000){2}{\rule{0.483pt}{0.400pt}}
\multiput(907.00,228.92)(1.142,-0.492){21}{\rule{1.000pt}{0.119pt}}
\multiput(907.00,229.17)(24.924,-12.000){2}{\rule{0.500pt}{0.400pt}}
\multiput(934.00,216.92)(1.203,-0.492){19}{\rule{1.045pt}{0.118pt}}
\multiput(934.00,217.17)(23.830,-11.000){2}{\rule{0.523pt}{0.400pt}}
\multiput(960.00,205.92)(1.381,-0.491){17}{\rule{1.180pt}{0.118pt}}
\multiput(960.00,206.17)(24.551,-10.000){2}{\rule{0.590pt}{0.400pt}}
\multiput(987.00,195.93)(1.485,-0.489){15}{\rule{1.256pt}{0.118pt}}
\multiput(987.00,196.17)(23.394,-9.000){2}{\rule{0.628pt}{0.400pt}}
\multiput(1013.00,186.93)(1.682,-0.488){13}{\rule{1.400pt}{0.117pt}}
\multiput(1013.00,187.17)(23.094,-8.000){2}{\rule{0.700pt}{0.400pt}}
\multiput(1039.00,178.93)(1.748,-0.488){13}{\rule{1.450pt}{0.117pt}}
\multiput(1039.00,179.17)(23.990,-8.000){2}{\rule{0.725pt}{0.400pt}}
\multiput(1066.00,170.93)(2.299,-0.482){9}{\rule{1.833pt}{0.116pt}}
\multiput(1066.00,171.17)(22.195,-6.000){2}{\rule{0.917pt}{0.400pt}}
\multiput(1092.00,164.93)(2.389,-0.482){9}{\rule{1.900pt}{0.116pt}}
\multiput(1092.00,165.17)(23.056,-6.000){2}{\rule{0.950pt}{0.400pt}}
\multiput(1119.00,158.93)(2.299,-0.482){9}{\rule{1.833pt}{0.116pt}}
\multiput(1119.00,159.17)(22.195,-6.000){2}{\rule{0.917pt}{0.400pt}}
\multiput(1145.00,152.94)(3.844,-0.468){5}{\rule{2.800pt}{0.113pt}}
\multiput(1145.00,153.17)(21.188,-4.000){2}{\rule{1.400pt}{0.400pt}}
\multiput(1172.00,148.94)(3.698,-0.468){5}{\rule{2.700pt}{0.113pt}}
\multiput(1172.00,149.17)(20.396,-4.000){2}{\rule{1.350pt}{0.400pt}}
\multiput(1198.00,144.94)(3.844,-0.468){5}{\rule{2.800pt}{0.113pt}}
\multiput(1198.00,145.17)(21.188,-4.000){2}{\rule{1.400pt}{0.400pt}}
\put(1225,140.17){\rule{5.300pt}{0.400pt}}
\multiput(1225.00,141.17)(15.000,-2.000){2}{\rule{2.650pt}{0.400pt}}
\multiput(1251.00,138.95)(5.597,-0.447){3}{\rule{3.567pt}{0.108pt}}
\multiput(1251.00,139.17)(18.597,-3.000){2}{\rule{1.783pt}{0.400pt}}
\put(1277,135.17){\rule{5.500pt}{0.400pt}}
\multiput(1277.00,136.17)(15.584,-2.000){2}{\rule{2.750pt}{0.400pt}}
\put(1304,133.17){\rule{5.300pt}{0.400pt}}
\multiput(1304.00,134.17)(15.000,-2.000){2}{\rule{2.650pt}{0.400pt}}
\put(1330,131.67){\rule{6.504pt}{0.400pt}}
\multiput(1330.00,132.17)(13.500,-1.000){2}{\rule{3.252pt}{0.400pt}}
\put(1357,130.17){\rule{5.300pt}{0.400pt}}
\multiput(1357.00,131.17)(15.000,-2.000){2}{\rule{2.650pt}{0.400pt}}
\put(1383,128.67){\rule{6.504pt}{0.400pt}}
\multiput(1383.00,129.17)(13.500,-1.000){2}{\rule{3.252pt}{0.400pt}}
\put(1410,127.67){\rule{6.263pt}{0.400pt}}
\multiput(1410.00,128.17)(13.000,-1.000){2}{\rule{3.132pt}{0.400pt}}
\put(1436,128){\usebox{\plotpoint}}
\put(1306,767){\makebox(0,0)[r]{$L=3$}}
\put(1350,767){\raisebox{-.8pt}{\makebox(0,0){$\Diamond$}}}
\put(266,803){\raisebox{-.8pt}{\makebox(0,0){$\Diamond$}}}
\put(448,604){\raisebox{-.8pt}{\makebox(0,0){$\Diamond$}}}
\put(630,415){\raisebox{-.8pt}{\makebox(0,0){$\Diamond$}}}
\put(811,273){\raisebox{-.8pt}{\makebox(0,0){$\Diamond$}}}
\put(993,191){\raisebox{-.8pt}{\makebox(0,0){$\Diamond$}}}
\put(1175,154){\raisebox{-.8pt}{\makebox(0,0){$\Diamond$}}}
\put(1356,138){\raisebox{-.8pt}{\makebox(0,0){$\Diamond$}}}
\sbox{\plotpoint}{\rule[-0.400pt]{0.800pt}{0.800pt}}%
\put(1306,722){\makebox(0,0)[r]{$L=4$}}
\put(1350,722){\makebox(0,0){$+$}}
\put(309,749){\makebox(0,0){$+$}}
\put(434,624){\makebox(0,0){$+$}}
\put(560,496){\makebox(0,0){$+$}}
\put(685,380){\makebox(0,0){$+$}}
\put(811,283){\makebox(0,0){$+$}}
\put(936,215){\makebox(0,0){$+$}}
\put(1062,170){\makebox(0,0){$+$}}
\put(1187,142){\makebox(0,0){$+$}}
\sbox{\plotpoint}{\rule[-0.500pt]{1.000pt}{1.000pt}}%
\put(1306,677){\makebox(0,0)[r]{$L=5$}}
\put(1350,677){\raisebox{-.8pt}{\makebox(0,0){$\Box$}}}
\put(420,600){\raisebox{-.8pt}{\makebox(0,0){$\Box$}}}
\put(581,455){\raisebox{-.8pt}{\makebox(0,0){$\Box$}}}
\put(742,331){\raisebox{-.8pt}{\makebox(0,0){$\Box$}}}
\put(904,234){\raisebox{-.8pt}{\makebox(0,0){$\Box$}}}
\put(1065,172){\raisebox{-.8pt}{\makebox(0,0){$\Box$}}}
\put(1226,141){\raisebox{-.8pt}{\makebox(0,0){$\Box$}}}
\put(1387,125){\raisebox{-.8pt}{\makebox(0,0){$\Box$}}}
\end{picture}

\caption{The  Binder  parameter  $g$  as  a function of the scaled reduced
temperature $(T-T_{c})L^{1/ \nu}$. The line is just guide to the eye.}
\label{fig:4}
\end{figure}
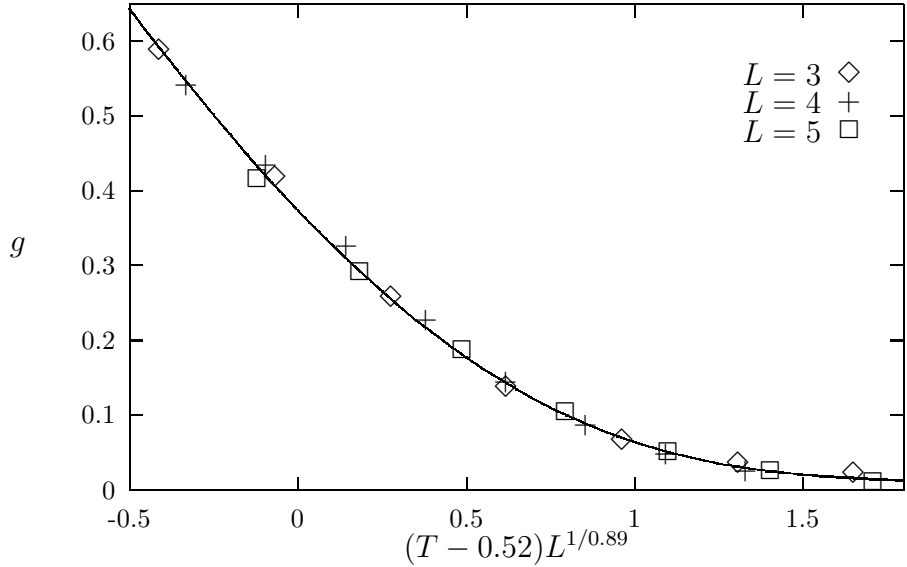

\end{document}